\documentclass[showpacs,preprintnumbers,pre]{revtex4}
\usepackage{graphicx}
\usepackage{dcolumn}
\usepackage{bm}
\usepackage{epsf}
\begin{document}
\title{
Microstructure and velocity of field-driven solid-on-solid interfaces
moving under stochastic dynamics with local energy barriers
}

\author{
G.~M.~Buend{\'{\i}}a$^{1,2}$}\email{buendia@usb.ve}
\author{
P.~A.~ Rikvold$^{2,3,4}$}\email{rikvold@csit.fsu.edu}
\author{M.~Kolesik$^{5,6}$}\email{kolesik@acms.arizona.edu}
\affiliation{
$^1$Department of Physics, Universidad Sim{\'o}n Bol{\'{\i}}var, Caracas 1080, Venezuela\\
$^2$School of Computational Science, Florida State University, Tallahassee, Florida
32306-4120, USA\\
$^3$Center for Materials Research and Technology
and Department of Physics,
Florida State University, Tallahassee, Florida 32306-4350, USA\\
$^4$ National High Magnetic Field Laboratory, Tallahassee, Florida 32310, USA\\
$^5$Institute of Physics, Slovak Academy of Sciences,
Bratislava, Slovak Republic\\
$^6$Optical Sciences Center, University of Arizona,
Tucson, Arizona 85721, USA\\
}
\date{\today}

\begin{abstract}
We study the microscopic structure and the stationary propagation velocity of
$(1+1)$-dimensional solid-on-solid interfaces in an Ising lattice-gas
model, which are driven far from
equilibrium by an applied force, such as a magnetic field or a
difference in (electro)chemical potential. We use an analytic
nonlinear-response approximation [P.~A.\ Rikvold and M.~Kolesik, J.\ Stat.\
Phys.\ {\bf 100}, 377 (2000)] together with
kinetic Monte Carlo simulations. Here we consider
interfaces that move under Arrhenius dynamics, which include a microscopic
energy barrier between the allowed Ising/lattice-gas states.
Two different dynamics are studied: the standard one-step
dynamics (OSD) [H.~C.\ Kang and W.~Weinberg, J.\ Chem.\ Phys.\ {\bf 90},
2824 (1992)] and the two-step transition-dynamics approximation (TDA) [T.\
Ala-Nissila, J.~Kjoll, and S.~C.\ Ying, Phys.\ Rev.\ B {\bf 46}, 846
(1992)]. In the OSD the effects of the applied force and the interaction
energies in the model factorize in the transition rates (soft dynamics),
while in the TDA such factorization is not possible
(hard dynamics). In full agreement with previous general theoretical results we
find that the local interface width under the TDA increases dramatically
with the applied force. In contrast, the interface structure with the OSD is
only weakly influenced by the force, in qualitative agreement with the
theoretical expectations. Results are also obtained for the
force-dependence and anisotropy of the interface velocity, which also
show differences in good agreement with the theoretical expectations for
the differences between soft and hard dynamics. Our results confirm that
different stochastic interface dynamics
that all obey detailed balance and the same conservation laws
nevertheless can lead to radically different interface
responses to an applied force.
\end{abstract}


\maketitle


\section{Introduction}
\label{sec:INTRO}

To understand the properties of materials one has to understand surfaces and interfaces: materials
interact with their environment through their surfaces, and material properties are profoundly influenced by internal interfaces. Processes related to surfaces and interfaces therefore play a critical role in nature and in a variety of technological applications, such as electronic, magnetic, and optical devices, sensors, catalysts, coatings, and many other industrial systems and processes.
\cite{DUKE02,ERTL97} Surfaces also play a vital role in biology and medicine, since most biological reactions occur at surfaces and interfaces. The applications of surface science in medicine range from the growth of bio-compatible surfaces for tissue cultures, through medical implants, to the design of innumerable medical devices.\cite{CAST02, TIRR02} It is therefore crucial to understand the fundamental processes occurring at surfaces and interfaces in order to provide means of controlling and manipulating such systems.
The large-scale properties of growing interfaces have been the object of an
enormous amount of work in recent years,\cite{BARA95,MEAK98} but much less attention has been paid to interfacial structure on a microscopic scale. This is unfortunate because 
many important interface properties, such as mobility and catalytic and chemical
activity, are largely determined by the microscopic interface structure. The microscopic
structure limits the interfacial propagation velocity under an external driving force,
such as an external field for a magnetic or dielectric domain wall, or
a difference in chemical potential between the bulk phases for 
a crystal surface.

Since the detailed microscopic mechanism of the interface motion is often not known, one 
standard way to gain insight in the process is by constructing a stochastic model that
mimics its essential features. Dynamics that conserve the order parameter, 
or ones that do not, must
be chosen according to the physical characteristics of the system. 
Once this is decided, there
are several dynamics in each category to choose among. 
It is well established that structures arising from different
dynamics that obey detailed balance and respect the same conservation
laws exhibit universal asymptotic large-scale features.
However, recent studies\cite{RIKV00B,RIKV02B,RIKV02,RIKV03}
show that there are important differences between
the {\it microstructures\/} of field-driven interfaces obtained with
different dynamics, and that these differences very significantly
influence important interface properties such as mobility.
A mean-field, nonlinear-response theory developed in
Refs.~\onlinecite{RIKV00B,RIKV02B} indicates that there are two different classes of nonconservative dynamics that lead to significantly different surface microstructures. These are the {\it soft} dynamics in which the single-site transition rates can be factorized into one part that depends only on the applied force and a second part that depends only on the interaction energies, and the
{ \it hard} dynamics for which this factorization 
is not possible.\cite{MARR99} By this
classification, the widely
used Glauber and Metropolis dynamics are hard.
Soft dynamics are 
appropriate for solidification or adsorption problems where the
driving force is a chemical-potential 
difference.\cite{GUO90,KOTR91}

In previous papers\cite{RIKV00B,RIKV02B} two of us introduced 
a dynamic mean-field approximation for the microstructure, based on the Burton-Cabrera-Frank 
 solid-on-solid (SOS) approximation.\cite{BURT51} For soft dynamics, interface
structures should remain independent of the applied force, while there should be a clear
dependence on the force for hard dynamics. Monte Carlo (MC) simulation results with the soft 
Glauber\cite{RIKV02B} and the standard hard Glauber dynamics\cite{RIKV02,RIKV02B,RIKV03} confirm these 
predictions. Moreover, SOS 
surfaces generated with the hard Glauber dynamics exhibit a skewness that is absent for the soft
Glauber dynamics. Since the mean-field results depend on the absence of short-range
 correlations along the interfaces, it remains an open question to which extent these characteristics are shared by all soft and hard dynamics.

For the present study we have considered two dynamics that include a local energy barrier
representing a transition state inserted between individual Ising or lattice-gas states.
Such Arrhenius dynamics, as they are often called, are appropriate in kinetic MC
 simulations of discrete Ising or lattice-gas models in which the discrete states serve as
approximations for high-probability configurations in an underlying continuous potential.\cite{MITC02,ALAN92A}
Examples are the study of diffusion in a lattice-gas model in a continuous corrugation 
potential,\cite{ALAN02} the relaxation process from the high-spin state of molecular bistable solids,\cite{GUTL94,REAL97,NISH03} or the approximation of a continuous spin model with strong uniaxial anisotropy by an
Ising model.\cite{EILO04,MUNO03} When applied to kinetic Ising lattice-gas models, Arrhenius dynamics give nucleation rates quite different from the ones given
by the standard Glauber dynamics;\cite{BUEN04A} another warning about the importance of choosing
the right dynamics for specific physical or chemical systems.

The two Arrhenius dynamics that we consider here are the common 
one-step-dynamics (OSD)\cite{KANG89, FICH91} and the two-step 
transition dynamic approximation (TDA).\cite{ALAN92,ALAN92A} 
The OSD dynamics are soft and the TDA dynamics are hard, 
according to the definitions given 
above. Their transition rates are defined in Sec. II. 
The OSD dynamics are commonly used in studies of adsorption, such as
the electrosorption of halides on single-crystal silver
electrodes.\cite{MITC00C,HAMA04,HAMA05} Among the experimental
systems that have been described with the TDA dynamics are the
diffusion of H atoms on single-crystal tungsten surfaces.\cite{ALAN92}

SOS surfaces belong to the Kardar-Parisi-Zhang 
(KPZ) dynamic universality class,\cite{BARA95,KARD86} in which 
the macroscopic, stationary distribution for flat,
moving interfaces is Gaussian, corresponding to a random walk with independent
increments. Nevertheless, the step heights in several discrete models in
this class are correlated at {\it short\/} distances.\cite{NEER97,KORN00C} 
In the mean-field approximation 
used here, these short-range correlations are 
ignored. The resulting discrepancies will be apparent when we compare the
theoretical results  
with kinetic MC simulations. 

The remainder of this paper is organized as follows. 
In Sec.~\ref{sec:MODEL} we introduce the SOS interface model and give the
transition rates for the TDA and OSD dynamics. 
In Sec.~\ref{sec:NLR} we summarize the mean-field approximation for the time 
evolution of the single-step probability density function (pdf), 
as well as its stationary form. We also give expressions for the class populations and
interface velocity in terms of the applied force, the temperature, and
the angle of the interface relative to 
the lattice axes.
In Sec.~\ref{sec:MC} we compare simulations and analytical predictions 
for the detailed stationary interface structure, including the asymmetry
of the simulated nonequilibrium interfaces. 
A summary and conclusion are provided in Sec.~\ref{sec:DISC}.

\section{Model and Dynamics}
\label{sec:MODEL}

The SOS interfaces are described by the nearest-neighbor 
$S=1/2$ Ising Hamiltonian with anisotropic, ferromagnetic interactions $J_x$ and $J_y$ in the $x$ and $y$ direction, 
respectively: 
\begin{equation}
{\cal H} = -\sum_{x,y} s_{x,y} \left( J_x s_{x+1,y} + J_y s_{x,y+1} 
+ H \right) 
\;, 
\label{eq:ham}
\end{equation}
where $s_{x,y}=\pm1$, $\sum_{x,y}$ runs over all sites, and 
 the applied field $H$ is the driving force. The interface is introduced by fixing 
$s_{x,y}=+1$ and $-$1 for large negative and positive $y$, respectively.  
Without loss of generality we take $H \ge 0$, such 
that the interface on average moves in the positive $y$ direction. 
This Ising model is equivalent to a lattice-gas model with local
occupation variables $c_{x,y} \in \{0,1\}$.\cite{YANG52B,PATH96}
Specifically, we identify $s=+1$ with $c=1$ (occupied or ``solid'') and
$s=-1$ with $c=0$ (empty or ``fluid''). The interactions in the 
Ising model, $J_{\alpha}$, are related to the ones in the lattice-gas model,
$\phi_{\alpha}$, as $J_{\alpha}=\phi_{\alpha}/4 $, and the applied 
field is related to
the lattice-gas chemical potential $\mu$ as $H = (\mu - \mu_0)/2$, where
$\mu_0 = -4(J_x+J_y)=-(\phi_{x}+\phi_{y})$ is the coexistence value of $\mu$. 
Here we will
use Ising or lattice-gas language interchangeably as we feel it makes a
particular aspect of the discussion clearer. 

\begin{figure}[ht] 
\includegraphics[angle=0,width=.50\textwidth]{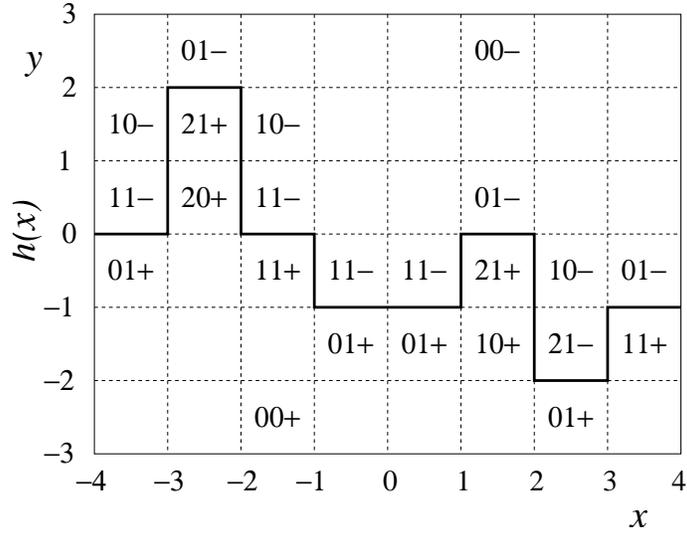} 
\caption[]{
A short segment of an SOS interface $y=h(x)$ between a positively
magnetized phase (or ``solid'' phase in the lattice-gas picture) 
below and a negative (or ``fluid'') phase
above. The step heights are $\delta(x) = h(x+1/2) - h(x-1/2)$. Interface
sites representative of the different SOS spin classes (see
Table~\protect\ref{table:class} 
and Table~\protect\ref{table:class2}) are marked with the
notation $jks$ explained in the text. Sites in the uniform bulk phases are
$00-$ and $00+$. This interface was randomly 
generated with a symmetric step-height 
distribution, corresponding to $\phi = 0$.
From Ref.~\protect\onlinecite{RIKV02B}. 
}
\label{fig:pict}
\end{figure}

The SOS model considers an interface in a lattice-gas or $S=1/2$
Ising system on a square lattice of unit lattice constant
as a single-valued integer function $h(x)$ of the $x$-coordinate, with
steps $\delta(x) = h(x+1/2) - h(x-1/2)$ at integer values of $x$.
A typical SOS interface configuration is shown in Fig.~\ref{fig:pict}. In
this paper the two possible states of the site
$(x,y)$ are denoted by the two ``spin'' values $s_{x,y} = \pm1$.
(In order that the step positions and the interface heights be
integer as stated above, we place the spins at odd
half-integer values of $x$ and $y$, i.e., at the centers of the
unit cells separated by dotted lines in Fig.~\ref{fig:pict}.)
                                                                          
The interface will be made to evolve under two different dynamics that contain a microscopic
energy barrier against individual spin flips. These are the transition 
dynamics 
approximation (TDA)\cite{ALAN92, ALAN92A} and the so-called one-step 
dynamics (OSD).\cite{KANG89, FICH91} The barrier represents a 
transition state which is inserted between the states
allowed in the Hamiltonian, such as a saddle point in a corrugation potential for particle diffusion,\cite{MITC02,ALAN92,ALAN02}
 or a high energy associated with a transitional spin state that is not along one of the two directions allowed by the Ising Hamiltonian.\cite{EILO04,MUNO03}

\begin{figure}[ht]
\includegraphics[angle=0,width=.50\textwidth]{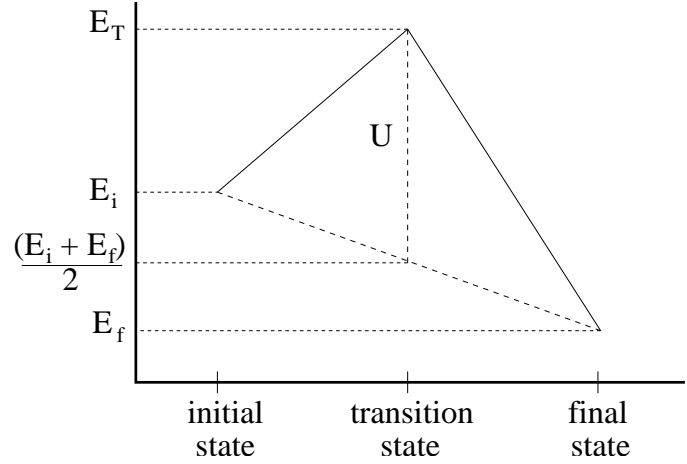}\\
\caption[]{Schematic picture of the transition barrier in
the symmetric Butler-Volmer approximation, used to calculate 
the TDA and OSD transition rates. 
After Ref.~\protect\onlinecite{BUEN04A}.
}
\label{fig:barrier}
\end{figure}

Here we express the transition-state energy by the following approximation,\cite{KANG89,FICH91,ALAN92,ALAN92A}

\begin{equation}
E_T=\frac{E_i + E_f}{2}+U ,
\label{eq:tener}
\end{equation}
where $E_i$ and $E_f$ are the initial and final energies, and $U$ is the bare, microscopic energy barrier. In electrochemical applications, such as electron- or ion-transfer reactions, this corresponds to the symmetric Butler-Volmer approximation.\cite{SCHM96} The construction 
corresponding to Eq.~(\ref{eq:tener}) is illustrated in Fig.~\ref{fig:barrier}.

Both the TDA and the OSD are single-spin-flip (nonconservative) dynamics that satisfy detailed 
balance.
This ensures the approach to
equilibrium, which in this case is a uniformly positive
phase with the interface pushed off to positive infinity.
Such dynamics are defined by a single-spin transition rate,
$W(s_{x,y} \rightarrow -s_{x,y}) = W(\beta \Delta E , \beta U)$. Here
$\beta$ is the inverse of the temperature $T$ (Boltzmann's constant is
taken as unity), $\Delta E = E_f - E_i$ 
is the energy change corresponding to a
successful spin flip, and $U$ determines the energy barrier
between the two states through Eq.~(\ref{eq:tener}).
 The detailed-balance condition (valid for transitions between
allowed states) is expressed as
$W(\beta \Delta E, \beta U) / W(- \beta \Delta E , \beta U)
= e^{- \beta \Delta E}$, where the right-hand side is independent of $U$.

The transition rates for the TDA and the OSD 
dynamics with the transition-state energy $E_T$ given 
by Eq.~(\ref{eq:tener}) are defined as\cite{ALAN92,ALAN92A,ALAN02}
\begin{equation}
W_{\rm {TDA}}=\frac{1}{1+\exp[\beta(E_T-E_i)]}\frac{1}{1+\exp[\beta(E_f-E_T)]}
\label{eq:wtda}
\end{equation}
and \cite{KANG89,FICH91}
\begin{equation}
W_{\rm {OSD}}=\exp[-\beta(E_T-E_i)]=\exp(-\beta U)\exp[-\beta \Delta E/2],
\label{eq:wosd}
\end{equation}
respectively.

Notice that the TDA transition rate cannot be factorized into one part that depends only on
the interaction energy and another that depends only on the applied field;
thus it belongs to the class of dynamics defined as hard.\cite{MARR99} 
The OSD dynamics can be factorized 
this way and thus is classified as soft.\cite{MARR99}  
Another important difference 
between the TDA and the OSD is that $W_{\rm TDA}$ is restricted to be between $0$ and $1$,
while there is no upper bound on $W_{\rm OSD}$. As a consequence, the same difference is 
observed for the propagation velocities. One may however question the
 physical realism of
cases in which the transition-state energy $E_T$ is below the initial 
energy $E_i$ so that $W_{\rm {OSD}}>1$. 
In order to preserve the
SOS configuration at all times, flips are allowed only at
sites which have exactly one broken bond in the $y$ direction.

With the Ising Hamiltonian there are only a finite 
number of different values of $\Delta E$. The spins can therefore be 
divided into classes,\cite{SPOH93,GILM76,BORT75,NOVO95A} 
labeled by the spin value $s$ and the number 
of broken bonds between the spin and its nearest neighbors in the $x$ and 
$y$ directions, $j$ and $k$, respectively. 
The ten spin classes consistent with the SOS model are denoted $jks$ 
with $j \in \{0,1,2\}$ and $k \in \{0,1\}$. They are shown in
Fig.~\ref{fig:pict} and listed in 
Table~\ref{table:class}. 

In the SOS model the heights of the individual 
steps are assumed to be statistically independent and 
identically distributed. The step-height probability density function (pdf) 
is given by the interaction energy corresponding to the $|\delta(x)|$ broken 
$J_x$-bonds between spins in the columns centered 
at $(x-1/2)$ and  $(x+1/2)$ as  
\begin{equation}
p[\delta(x)] = Z(\phi)^{-1} X^{|\delta(x)|}
\ e^{ \gamma(\phi) \delta(x) } \;. 
\label{eq:step_pdf}
\end{equation}
The factor $X$ determines the width of the pdf, 
and $\gamma(\phi)$ is a Lagrange multiplier which maintains the mean step 
height at an $x$-independent value, $\langle \delta(x) \rangle = \tan \phi$, 
where $\phi$ is the overall angle between the interface and the $x$ axis. 
In equilibrium, $X$ is simply the Boltzmann 
factor, $e^{- 2 \beta J_x}$.
The partition function is, 
\begin{equation}
Z(\phi)
=
\sum_{\delta = -\infty}^{+\infty} X^{|\delta|} e^{ \gamma(\phi) \delta } 
= 
\frac{1-X^2}{1 - 2 X \cosh \gamma(\phi) + X^2} ,
\label{eq:Z}
\end{equation}
where $\gamma(\phi)$ is given by 
\begin{equation}
e^{\gamma (\phi)} 
= 
\frac{ \left(1+X^2 \right)\tan \phi 
+ \left[ \left( 1 - X^2 \right)^2 \tan^2 \phi + 4 X^2 \right]^{1/2}}
{2 X \left( 1 + \tan \phi \right)} 
\label{eq:chgam}
\end{equation}
(see details in Refs.~\onlinecite{RIKV00B,RIKV02B}). 
Simple results are obtained for 
$\phi = 0$, which yields $\gamma(0) = 0$ and 
\begin{equation}
Z(0) = (1+X)/(1-X) \;,
\label{eq:Z0}
\end{equation} 
and for $\phi=45^{\circ}$ (see Ref.~\onlinecite{RIKV02B}).

The mean spin-class 
populations, $\langle n(jks) \rangle$, are all obtained from the 
product of the independent pdfs for $\delta(x)$ and $\delta(x$+1). 
Symmetry of $p[\delta(x)]$ under the transformation 
$(x,\phi,\delta) \rightarrow (-x,-\phi,-\delta)$ ensures that 
$\langle n(jk-) \rangle = \langle n(jk+) \rangle$ for all $j$ and $k$. 
Numerical results illustrating the breakdown of this up/down symmetry for 
large $H$ are discussed in Sec.~\ref{sec:MC}. The general expressions for
the class populations are given in the first column of Table~\ref{table:class2}; details
of the calculation can be found in Ref.~\onlinecite{RIKV00B}. The
 results for each of the dynamics are obtained by substituting their
respective values of $X$, which will be calculated in the next Section. 

Whenever a spin at the interface flips from $-$1 to +1, 
the corresponding column of the interface advances by one 
lattice constant in the $y$ direction. Conversely, a column 
(not necessarily the same one) recedes by one lattice constant 
when a spin at the interface flips from +1 to $-$1. 
The mean velocity of the interface in the $y$-direction, 
$\langle v_y \rangle$, is the difference between the 
rates of forward and backward steps, averaged over the whole
interface.  The energy changes corresponding to the flips are 
given in the third column in Table~\ref{table:class}. 
The sum over $x$ that gives the average velocity can be rearranged
into a sum over the spin classes of transition rates, weighted by the
average class populations. Since the spin-class populations on both 
sides of the interface are equal in this approximation, the contribution 
to $\langle v_y \rangle$  from sites in the classes $jk-$ and $jk+$ 
therefore becomes 
\begin{equation}
\langle v_y(jk) \rangle 
= 
W \left( \beta \Delta E(jk-) ,\beta U  \right)
-
W \left( \beta \Delta E(jk+) , \beta U \right) 
 \;. 
\label{eq:generalv}
\end{equation}
The mean propagation velocity perpendicular to the interface 
becomes  
\begin{equation}
\langle v_\perp (T,H,\phi) \rangle 
= 
\cos ( \phi ) \langle v_y \rangle 
= 
\cos ( \phi ) \sum_{j,k} \langle n(jks) \rangle \langle v_y (jk) \rangle 
\;, 
\label{eq:totalv}
\end{equation}
where the sum runs over the classes included in Table~\ref{table:class2}. 
It has been shown in Ref.~\onlinecite{RIKV00B} 
that Eq.~(\ref{eq:totalv}) reduces to the results for 
the single-step\cite{DEVI92,SPOH93,MEAK86,PLIS87} and the polynuclear 
growth\cite{DEVI92,KRUG89,KERT89}
models at low temperatures for large and small $\phi$, respectively. 

\section{Nonlinear Response} 
\label{sec:NLR} 

With $X = e^{-2 \beta J_x}$, i.e, independent of $H$, the results in Table~\ref{table:class2} 
correspond to a linear-response approximation for the velocity. 
 In previous papers\cite{RIKV00B,RIKV02B} an expression for a field-dependent $X(T,H)$ was
obtained, based on a dynamic mean-field approximation for the equation of motion for the
single-step pdf together with a detailed-balance argument for the stationary state. This improved
non-linear response approximation gives (see Ref.~\onlinecite{RIKV02B} for details of the calculation),
\begin{equation}
X(T,H) = e^{-2 \beta J_x} 
\left\{
\frac{e^{-2 \beta H}W[\beta(-2H-4J_x),\beta U] + e^{2\beta H}W[\beta(2H-4J_x),\beta U]}
{W[\beta(-2H-4J_x),\beta U] + W[\beta(2H-4J_x),\beta U]}
\right\}^{1/2}
\;,
\label{eq:XTH}
\end{equation}
which is independent of $\gamma(\phi)$. Here $W(\beta \Delta E,\beta U)$ 
are the transition 
rates associated with the reversal of a single spin. The values
 of $\Delta E$, $E_T - E_i$, and $E_f - E_T$ are  given in
Table~\ref{table:class} for the different spin classes.

Equation~(\ref{eq:XTH}) shows that $X(T,H)$ depends on the specific 
dynamics, except for $H=0$, where it reduces to its equilibrium value, 
$X(T,0) = e^{-2 \beta J_x}$.
It is easy to see from the equation that for soft dynamics, where the field and 
interaction contributions to the transition rates factorize, the $H$-dependence in Eq.~(\ref{eq:XTH})
cancels out. 
In Ref.~\onlinecite{RIKV02} it was demonstrated that the soft Glauber 
dynamics yields an SOS interface 
that is identical to the equilibrium SOS interface at $H=0$ and the same 
temperature, regardless of the value of $H$. 
Hard dynamics, such as the standard Glauber and Metropolis dynamics and the TDA, lead to a 
nontrivial field dependence in $X$. 

Inserting the transition rates corresponding to the TDA and the OSD dynamics, 
defined by Eq.~(\ref{eq:wtda}) and  Eq.~(\ref{eq:wosd}), respectively, into Eq.~(\ref{eq:XTH}), we explicitly get 
\begin{equation}
X_{\rm TDA}(T,H) = 
e^{-2 \beta J_x} 
\left\{
\frac{e^{2 \beta J_x} \cosh(2 \beta H) + e^{-2\beta J_x} 
+ 2 \cosh(\beta U) \cosh(\beta H)}
{e^{-2 \beta J_x} \cosh(2 \beta H) + e^{2\beta J_x}
+ 2 \cosh(\beta U) \cosh(\beta H)}
 \right\}^{1/2}
\;
\label{eq:XTDA}
\end{equation}
and
\begin{equation}
X_{\rm OSD}(T,H) =
e^{-2 \beta J_x}=X(T,0).
\,
\label{eq:XOSD}
\end{equation}
We note that $X_{\rm TDA}$ is similar, but not identical, to the one for the standard
Glauber dynamics, Eq.~(18) of Ref.~\onlinecite{RIKV02B}.
The spin-class populations 
  listed in Table~\ref{table:class2} 
can now be calculated explicitly for each of the dynamics by replacing $X$ with its
corresponding value. The expressions for the contributions to the mean velocity in the $y$ direction,
Eq.~(\ref{eq:generalv}), for
each class in the TDA and the OSD dynamics are given in the third and fourth columns of
 Table~\ref{table:class2}, respectively. 

In the next Section we show that the nonlinear-response approximation 
gives good agreement with MC simulations of driven, flat SOS interfaces 
evolving under the TDA and OSD dynamics for a wide range of fields and temperatures. 

\section{Comparison with Monte Carlo Simulations} 
\label{sec:MC} 

We calculated the step-height distributions, 
propagation velocities, 
and spin-class populations, analytically and by kinetic MC simulations,
 for both the TDA and the OSD dynamics in the isotropic case, $J_x = J_y = J$. 
 The details of our 
particular implementation 
of the $n$-fold way rejection-free MC 
algorithm\cite{BORT75, GILM76} are essentially the same as described in Ref.~\onlinecite{RIKV00B}, 
except for two points. The first is that only transitions from the classes with
 one broken $y$-bond 
($k=1$) are allowed, so as to preserve the SOS interface structure. The second difference
is that the present code uses {\it continuous time}\cite{KORN99} to accommodate the large transition
rates that are possible with the OSD dynamics.
By keeping only the interface sites in memory, the algorithm is not subject 
to any size restriction in the $y$ direction, and simulations can be 
carried out for arbitrarily long times. 

The numerical results presented here are based on MC simulations mostly at the two 
temperatures, $T = 0.2T_c$ and~0.6$T_c$ ($T_c = -2J/\ln(\sqrt{2} -1)
\approx 2.269J$ is the critical temperature for the isotropic,
square-lattice Ising model\cite{ONSA44}), with  
$L_x = 10\,000$ and fixed $\phi$ between 0 and $45^\circ$. The microscopic transition
barrier, $U$ (see Eq.(2)), is chosen to be $0.5J$. This is the 
same value used in
a previous study of nucleation with the OSD and TDA dynamics.\cite{BUEN04A} From Eq.~(\ref{eq:wosd}) it is clear
that for the OSD dynamics $U$ only appears in a temperature-dependent scaling
factor in the transition rate. It thus has no influence on the interface 
structure. For the TDA dynamics, on the other hand, an increase in
$U$ leads to a decrease in the local interface width and
consequently in the propagation velocity. Also, the observed skewness
increases, suggesting increasing short-range correlations between
the step heights. These results are discussed in detail in a separate 
paper.\cite{BUEN06} 

In order to ensure stationarity we ran the simulation for 50\,000 
$n$-fold way updates per updatable spin (UPS)  
before taking any measurements. 
 Stationary 
class populations and interface velocities were averaged over 50\,000~UPS. 
For the stronger fields at $T=0.2T_c$ we used ten times as many UPS. Adequate statistics for one- and two-step pdfs were ensured by the large $L_x$. 

\subsection{Stationary single-step probability densities}
\label{sec:MCss}

\begin{figure}[ht]
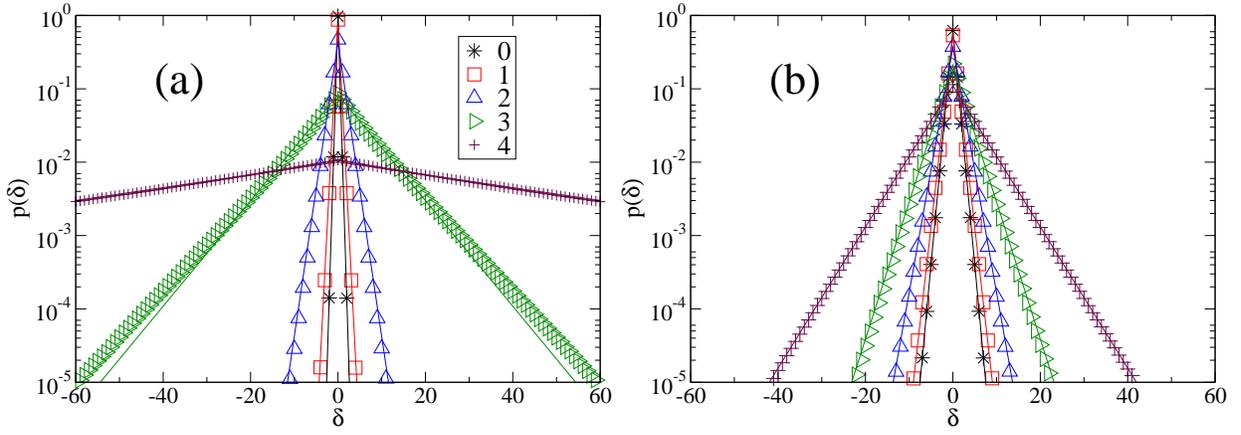
 
\includegraphics[angle=0,width=.45\textwidth]{pd_TDA_a.eps}
\includegraphics[angle=0,width=.45\textwidth]{pd_TDA_b.eps}
\caption[]{
(Color online) MC (data points) and analytical (solid lines) results for the stationary 
single-step pdf calculated with the TDA dynamics, shown on a logarithmic
 scale vs $\delta$, for the values of $H/J$ given in the legend.
(a) $T=0.2T_c$. 
(b) $T=0.6T_c$. The symbols (and colors)  have
the same interpretations in (a) and (b).
}
\label{fig:pd_TDA}
\end{figure}

\begin{figure}[ht]
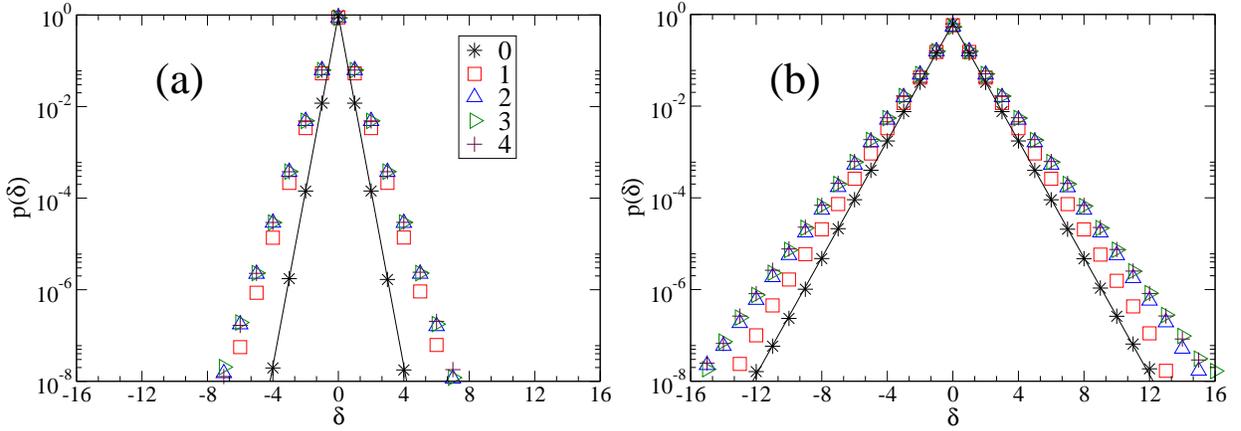

\includegraphics[angle=0,width=.45\textwidth]{pd_1SD_a.eps}
\includegraphics[angle=0,width=.45\textwidth]{pd_1SD_b.eps}
\caption[]{
(Color online) The stationary
single-step pdf calculated with the OSD dynamics, shown on a logarithmic scale vs $\delta$.
The data points indicate MC results and the straight lines are the theoretical predicted
values (independent of $H$). The values of $H/J$ are given in the legend.
(a) $T=0.2T_c$.
(b) $T=0.6T_c$. The symbols and colors have
the same interpretations in (a) and (b).
 Note the very different scales from Fig.~\ref{fig:pd_TDA}.
}
\label{fig:pd_1SD}
\end{figure}

Stationary single-step pdfs were obtained by MC simulation at $T=0.2T_c$ and 
$0.6T_c$ for $\phi = 0$ and 
several values of $H$. The simulation data
and the theoretical results for $p[\delta]$ are shown in Fig.~\ref{fig:pd_TDA} and
Fig.~\ref{fig:pd_1SD} 
for the TDA and
the OSD dynamics, respectively. The theoretical results are calculated
with 
Eq.~(\ref{eq:step_pdf}), with $X(T,H)$ from Eq.~(\ref{eq:XTDA}) 
for the TDA dynamics and from 
Eq.~(\ref{eq:XOSD}) for the OSD dynamics. 
For the TDA
dynamics, Fig.~\ref{fig:pd_TDA}, the agreement is excellent at the higher temperature for all the values of $H$ analyzed (up to $H/J=4$). However, at the lower temperature the agreement is
not very good for fields above $H/J=2.5$. For the OSD dynamics, Fig.~\ref{fig:pd_1SD} shows
that, contrary to the theoretical mean-field results, $p(\delta)$ depends somewhat on $H$. This dependence
is stronger for small fields and at the lower temperature.
 However, although not absent as expected from the mean-field approximation, the 
field  dependence is {\it much weaker} than for the TDA dynamics.

\begin{figure}[ht]
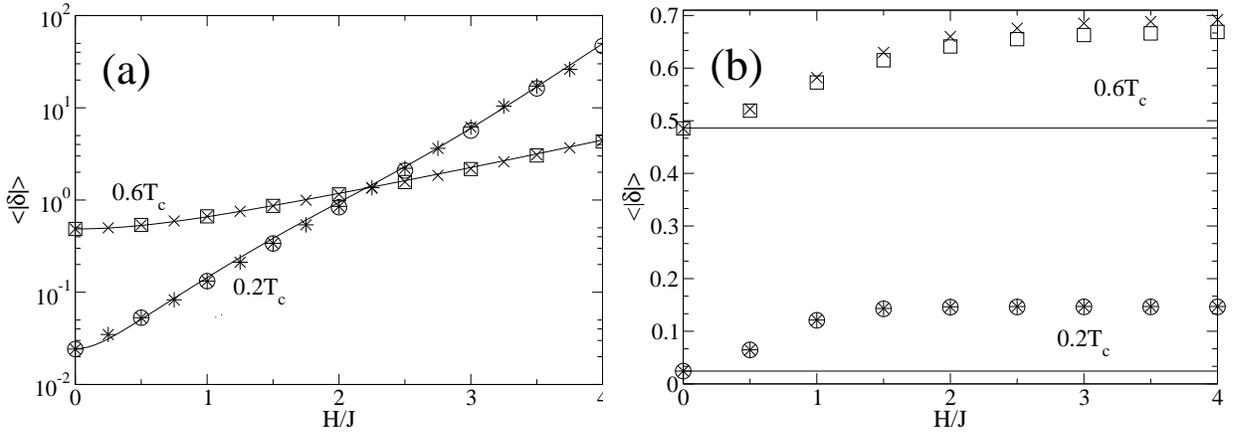
 
\includegraphics[angle=0, width=.45\textwidth]{dh_TDA.eps}
\includegraphics[angle=0,width=.45\textwidth]{dh_OSD.eps}
\caption[]{
Average stationary step height 
$\langle | \delta | \rangle$ 
vs $H$ for $\phi$=0 at $T$=$0.2T_c$ and~0.6$T_c$.
The curves represent the theoretical results. 
The MC data were obtained directly by summation 
over the simulated single-step pdfs (asterisks and crosses) and from the probability 
of zero step height (circles and squares). 
See the text for details. 
Curve with circles and asterisks: $T=0.2T_c$. 
Curve with squares and crosses : $T=0.6T_c$. 
(a) TDA dynamics, shown on a logarithmic vertical scale.
(b) OSD dynamics, shown on a linear vertical scale. In this and all the following figures, the statistical uncertainty is
much smaller than the symbol size.
}
\label{fig:dh}
\end{figure}

Another way to compare the analytical and simulation results is by calculating
$\langle | \delta | \rangle$ by summation of Eq.~(\ref{eq:step_pdf}),
$\langle | \delta | \rangle = 2X/\left( 1-X^2 \right)$,
with $X$ from Eq.~(\ref{eq:XTDA}) for the TDA dynamics or $X$ from Eq.~(\ref{eq:XOSD}) 
(independent of $H$) for the 
OSD, and comparing these values with the simulated ones. The simulation values for
$\langle | \delta | \rangle$ can be 
obtained in two ways: directly by summation over the
numerically obtained pdf, and also from the probability
of zero step height as
$\langle | \delta | \rangle = \left\{ p[0]^{-1} - p[0] \right\}/2$.
This latter expression is obtained by observing that 
$p[0] = (1-X)/(1+X)$ for $\phi = 0$, then solving this for $X$, 
and inserting the result in the above
equation for $\langle | \delta | \rangle$ in terms of $X$. 
The results are shown in 
Fig.~\ref{fig:dh} for both dynamics, for $\phi = 0$ at $T = 0.2T_c$ and~0.6$T_c$ calculated theoretically (solid lines) and
by MC simulation (symbols). 
The agreement between the simulation and theoretical
results for the TDA is excellent except at the low temperature, where for intermediate fields
a slight deviation can be seen. The results are similar
to those obtained with the standard Glauber dynamics. (See Fig.~5(a) of Ref.~\onlinecite{RIKV02B}.) Again the results for the OSD dynamics
show that, contrary to the theoretical prediction, there is a clear, albeit weak, dependence of
 the step height on the field.
The theoretical and simulation results for the OSD dynamics only coincide at $H/J=0$. 

The difference between the two dynamics is evident: 
the step heights for the OSD 
dynamics are weakly dependent on $H$, particularly for low values of $H$,
in contrast with the very strong $H$ dependence obtained for the
TDA dynamics. This behavior is typical for differences expected 
between soft and hard dynamics.\cite{RIKV02}

\subsection{Stationary interface velocities}
\label{sec:MCi}

\begin{figure}[ht]
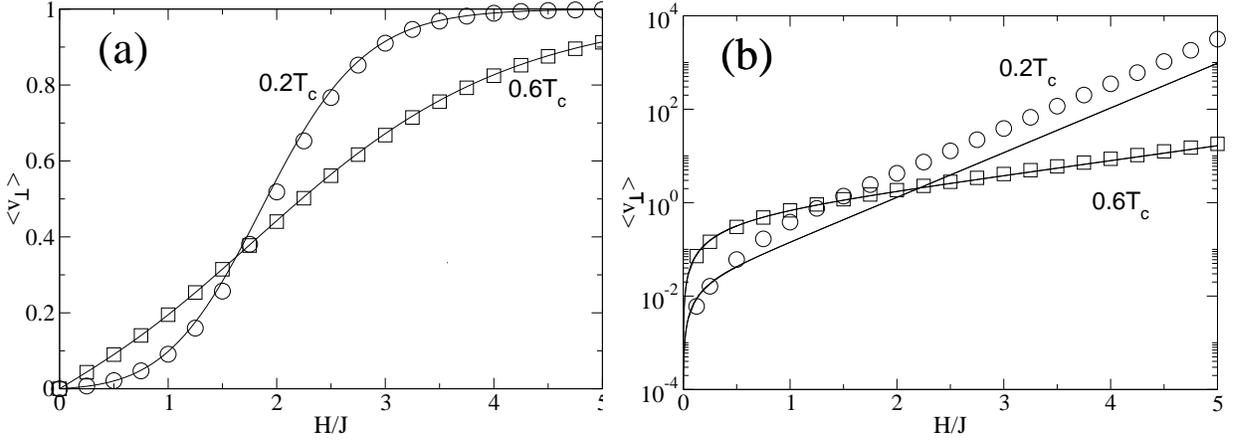
 
\includegraphics[angle=0,width=.45\textwidth]{vph_TDA.eps}
\includegraphics[angle=0,width=.45\textwidth]{vph_1SD.eps}
\caption[]{
The average stationary 
normal interface velocity $\langle v_\perp \rangle$ vs $H$ for 
$\phi = 0$. The MC results are shown as data points, circles for $T=0.2T_c$ and
squares for $T=0.6T_c$,   
and the theoretical results as solid curves. 
(a) TDA, shown on a linear vertical scale.
(b) OSD,
shown on a logarithmic vertical scale. 
}
\label{fig:vph}
\end{figure}

In this Section we compare the interface velocities 
obtained for the TDA and the OSD dynamics.
 In each case the velocities are
calculated with the analytical approximation, Eq.~(\ref{eq:totalv}), and by simulations.
Fig.~\ref{fig:vph}(a) and Fig.~\ref{fig:vph}(b) show the normal velocity vs $H$ for 
$\phi = 0$ for the TDA and the OSD dynamics, respectively.  
There is excellent agreement between the 
MC results and the nonlinear-response theory for the TDA dynamics, 
except for a 
slight disagreement seen at $0.2T_c$ between $H/J = 1.5$ and $2.5$.
The results are very similar to
 those obtained with the standard Glauber dynamics. (See Fig.~6 of Ref.~\onlinecite{RIKV02B}.)
However, for the OSD dynamics at the lower temperature,
the nonlinear-response
approximation underestimates the velocity, especially at 
higher fields. One of the main differences between the two dynamics is clearly seen in
Fig.~\ref{fig:vph}: the velocity is bounded by unity for the TDA while it increases exponentially
with $H$ for the OSD.

\begin{figure}[ht]
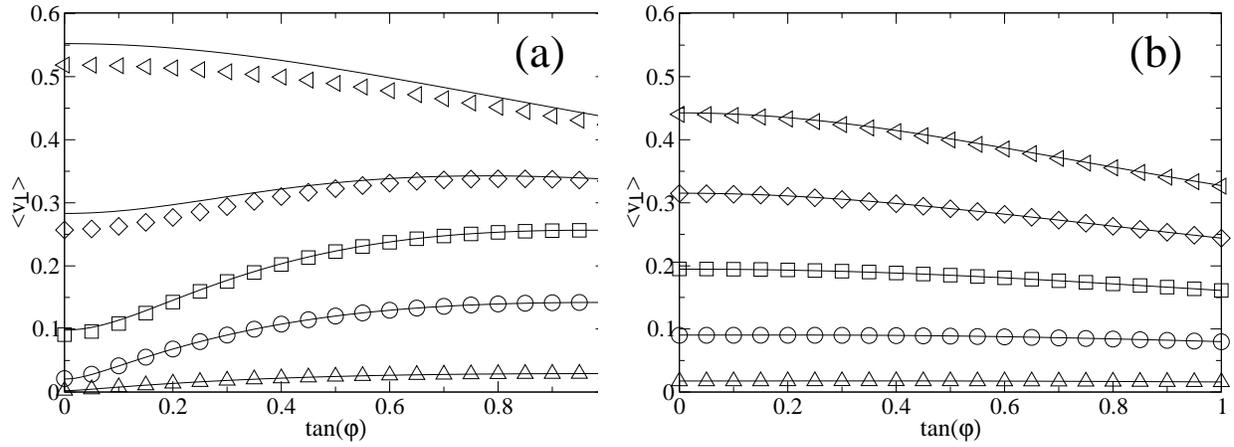

\includegraphics[width=.45\textwidth,angle=0]{vf_TDA_a.eps}
\includegraphics[width=.45\textwidth,angle=0]{vf_TDA_b.eps}
\caption[]{
The average stationary
normal interface velocity $\langle v_\perp \rangle$ vs $\tan \phi$, calculated with
the TDA dynamics,
for $H/J=0.1$ (up triangles), $0.5$ (circles), $1$ (squares), 
$1.5$ (diamonds), and $2.0$ (left triangles). The symbols represent MC data, and the solid curves analytical results.
(a) $T=0.2T_c$. (b) $T=0.6T_c$.
}
\label{fig:vang_TDA}
\end{figure}

\begin{figure}[ht]
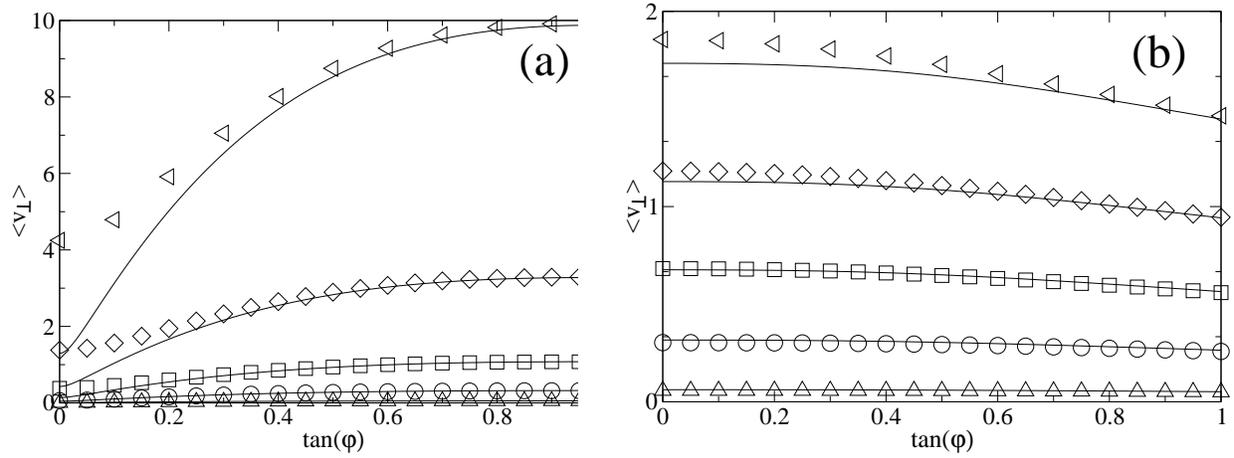

\includegraphics[width=.45\textwidth,angle=0]{vf_1SD_a.eps}
\includegraphics[width=.45\textwidth,angle=0]{vf_1SD_b.eps}
\caption[]{
The average stationary
normal interface velocity $\langle v_\perp \rangle$ vs $\tan \phi$, calculated with the OSD
dynamics,
for $H/J=0.1$ (up triangles), $0.5$ (circles), $1$ (squares), 
$1.5$ (diamonds), and $2.0$ (left triangles)
. The symbols represent MC data, and the solid curves analytical results.
(a) $T=0.2T_c$. (b) $T=0.6T_c$. Note the difference in the vertical scales.
}
\label{fig:vang_OSD}
\end{figure}

The dependence of the normal velocity on the tilt angle $\phi$ is shown 
 in
Fig.~\ref{fig:vang_TDA} and Fig.~\ref{fig:vang_OSD}, for the TDA and the OSD dynamics,
respectively. We show results for several values of $H/J$ at $T=0.2T_c$ and
$T=0.6T_c$. For the TDA dynamics, the agreement between the theoretical results and the 
simulations is excellent. The results are qualitatively similar to those obtained with
the standard Glauber dynamics. (See Fig.~7 of Ref.~\onlinecite{RIKV02B}.) For the
 OSD, the agreement between theory and simulation is excellent at the higher temperature. However, at the lower temperature and higher fields the agreement is only good at large values of $\phi$.

In both cases it is seen that at $T=0.2T_c$ in weak fields the velocity increases with 
$\phi$, in agreement with the polynuclear growth model at 
small angles and the single-step model for larger angles. However, for strong fields
the TDA dynamics change gradually to the reverse anisotropy of Eden-type
 models.\cite{MEAK86B,HIRS86} No such change is observed for the OSD dynamic. 
 At $T=0.6T_c$, on the other hand, both models behave very similarly. The velocity is 
nearly isotropic for weaker fields, while becoming Eden-like for stronger
fields.

\begin{figure}[ht]
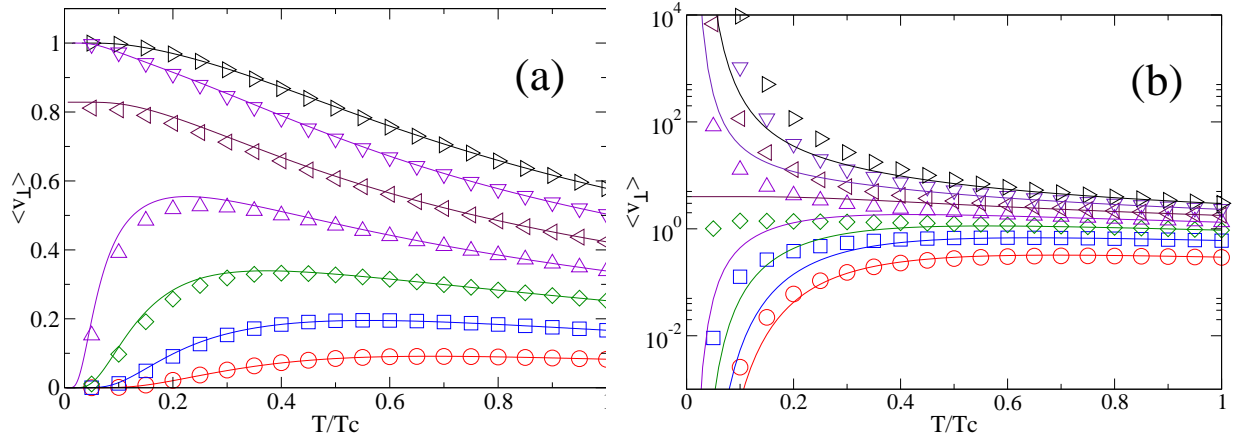

\includegraphics[angle=0,width=.45\textwidth]{vt_TDA.eps}
\includegraphics[angle=0,width=.45\textwidth]{vt_1SD.eps}
\caption[]{
(Color online) The average stationary
normal interface velocity $\langle v_\perp \rangle$ vs $T$
for $\phi = 0$ and $H/J$ between $0.5$ and $3.5$. MC data are represented
by data points
and analytical results by solid curves. 
From below to above, the values of $H/J$ are
0.5 (circles), 1.0 (squares), 1.5 (diamonds), 2.0 (up triangles), 2.5 (left triangles),
 3.0 (down triangles), and 3.5 (right triangles). Online, the colors of the curves and symbols match. 
(a) TDA, on a linear vertical scale. 
(b) OSD, on a logarithmic vertical scale.
}
\label{fig:vt}
\end{figure}

The temperature dependence of the normal interface velocity is shown in 
Fig.~\ref{fig:vt} for several values of $H/J$. 
For the TDA dynamics, Fig.~\ref{fig:vt}(a), the agreement between the simulations and the analytical results is excellent almost everywhere, except for a small discrepancy at
intermediate $T$ and $H$. This discrepancy is  expected from the 
results shown in Fig.~\ref{fig:vph}.
This figure shows that at low $T$, the
velocity changes steeply from zero to unity at some value
between $H/J=2$ and $H/J=2.5$,
developing a step 
discontinuity in $H$ at $T=0$. The results are qualitatively similar to
those obtained with
the standard Glauber dynamics. (See Fig.~8 of Ref.~\onlinecite{RIKV02B}.) For the OSD, Fig.~\ref{fig:vt}(b),
the theoretical and simulation results agree only when the temperature is higher than 
a minimum value that increases as the field increases. In this case the velocity
also has a step discontinuity at $T=0$:
for $H/J$ below $1.5$ the
velocity goes to zero. For stronger
fields the velocity at $T=0$ increases dramatically with $H$ and decreases very rapidly as the 
temperature increases. 
As can be seen from Eq.~(\ref{eq:totalv}) and
 Table~\protect\ref{table:class2}, 
the mean-field theory predicts that the contributions to the propagation 
velocity from each of the classes at low $T$ is proportional to 
$\exp \left[ - \beta \left( 2 J_x + U - H \right) \right]$. 
Thus, there is a discontinuity at $T=0$ for $H=2J_x+U$, i.e., for $H/J
=2.5$ for our selection of $U=0.5J$. Beyond this value of $H$, the velocity diverges to
infinity as $T\rightarrow 0$. Such a divergence at $T=0$ is not present for
the soft Glauber dynamics, where the velocity vanishes at $T=0$ for all values of $H$.\cite{RIKV03}

\subsection{Spin-class populations and skewness}
\label{sec:MCc}

\begin{figure}[ht]
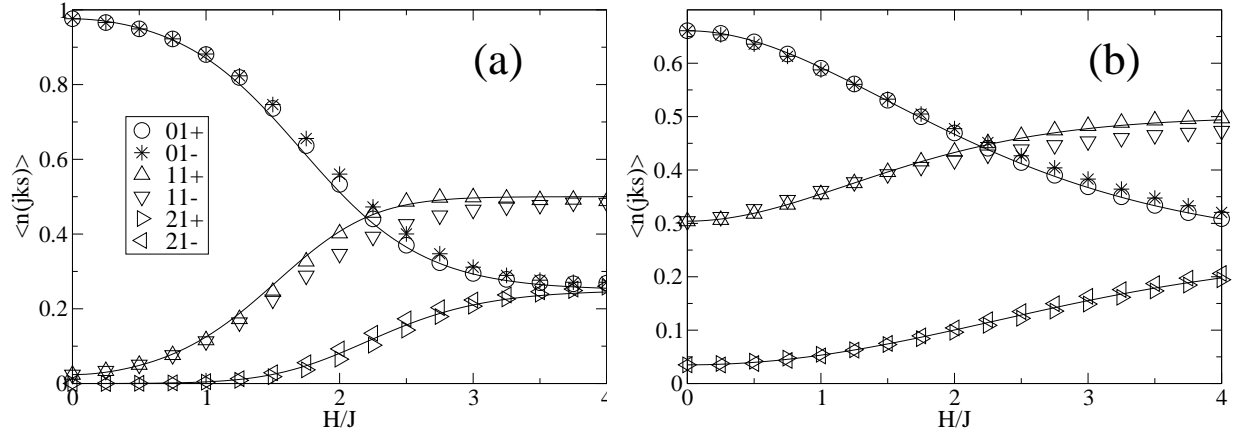
 
\includegraphics[angle=0,width=.45\textwidth]{pop_TDA_a.eps}
\includegraphics[angle=0,width=.45\textwidth]{pop_TDA_b.eps}
\caption[]{
Mean stationary class populations $\langle n(jks) \rangle$ vs $H/J$ 
for $\phi = 0$, calculated for the TDA dynamics. 
The simulation results are indicated by symbols, and the analytic 
approximations by solid curves. (a) $T=0.2T_c$. (b) $T=0.6T_c$. The symbols have the same interpretations in (a)
and (b).
Note the different vertical scales in the two parts.
}
\label{fig:pop_TDA}
\end{figure}

\begin{figure}[ht]
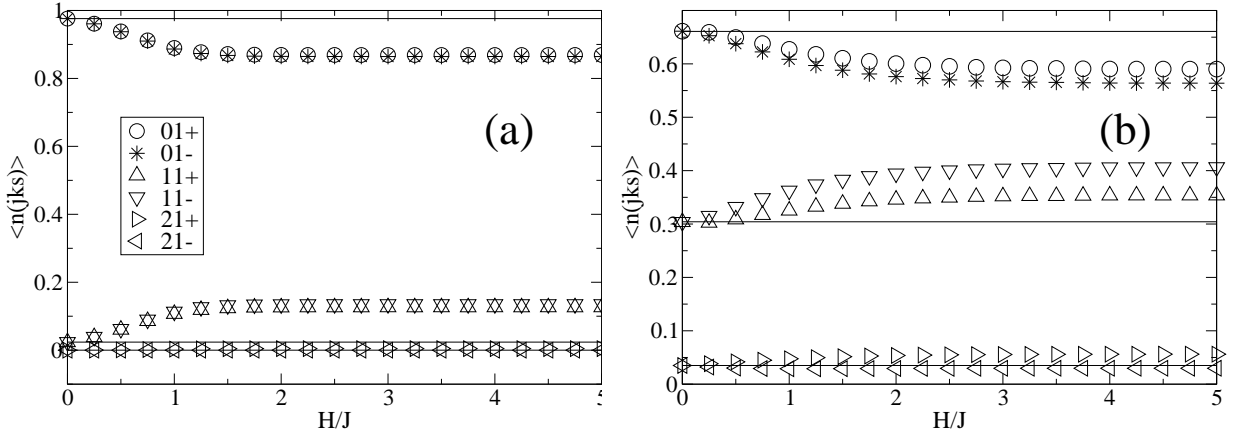

\includegraphics[angle=0,width=.45\textwidth]{pop_1SD_a.eps}
\includegraphics[angle=0,width=.45\textwidth]{pop_1SD_b.eps}
\caption[]{
Mean stationary class populations $\langle n(jks) \rangle$ vs $H/J$
for $\phi = 0$ calculated for the OSD dynamics.
The simulation results are indicated by the symbols, and the straight lines
indicate the theoretical predicted values (independent of $H$).
(a) $T=0.2T_c$. (b) $T=0.6T_c$. The symbols have the same interpretations in
(a) and (b). Note the different vertical scales in the two parts.
}
\label{fig:pop_1SD}
\end{figure}

Since the analytical predictions 
for the class populations are 
based on the assumption that different steps are statistically independent, 
a comparison with the simulation results 
gives a way of testing this assumption. The six mean class populations -- $\langle n(01s) \rangle$, 
$\langle n(11s) \rangle$, 
and $\langle n(21s) \rangle$ with $s = \pm 1$ -- for $\phi = 0$ at $T=0.2T_c$ and $0.6T_c$ are shown vs $H$ 
in Fig.~\ref{fig:pop_TDA} and Fig.~\ref{fig:pop_1SD} for the TDA and the OSD
dynamics, respectively. For the TDA at both 
temperatures, Fig.~\ref{fig:pop_TDA}, the 
analytical approximations follow the average of the populations 
for $s=+1$ and $s=-1$ quite well. However, at intermediate fields 
the simulations show that the population in front of the surface
($s = -1$) is quite different from the one behind it ($s=+1$). The 
 mean-field approximation seems to reproduce better the population behind
the surface. The results are qualitatively similar to those
obtained with
the standard Glauber dynamics. (See Fig.~9 of Ref.~\onlinecite{RIKV02B}.) 
 For the OSD, Fig.~\ref{fig:pop_1SD}, the
mean-field approximation predicts that the mean class populations should be independent of $H$ (since $X$ is independent of $H$), while
the simulation indicates a weak $H$-dependence. For small fields there is a clear dependence
of the population on the field, but as the field increases, the populations tend to fixed values
independent of the field. 
The $H$-dependence is consistent with the results 
for $\langle | \delta | \rangle$ and $X(T,H)$,
shown in Fig.~\ref{fig:dh}(b).

\begin{figure}[ht]
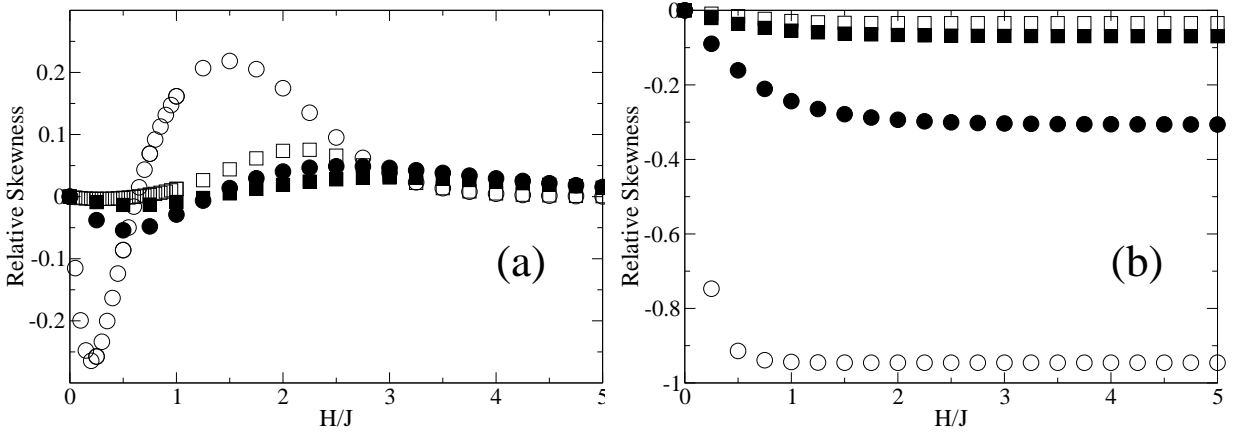
 
\includegraphics[angle=0,width=.45\textwidth]{skew_TDA.eps}
\includegraphics[angle=0,width=.45\textwidth]{skew_1SD.eps}
\caption[]{
The two relative skewness parameters $\rho$ (circles)  and 
$\epsilon$ (squares), defined in 
Eqs.~(\protect\ref{eq:rho}) and~(\protect\ref{eq:epsi}), respectively. 
The parameters are shown vs $H$ 
for $\phi = 0$, at $T=0.2T_c$ (empty symbols) and at $T=0.6T_c$ (filled symbols).
(a) TDA. (b) OSD. 
Note the different vertical scales in the two parts. 
}
\label{fig:skew}
\end{figure}

The short-range correlations between neighboring steps are responsible for the skewness between the spin populations on the leading and trailing edges 
of the interface that appears in the simulation results. This phenomenon is commonly observed in driven interfaces. 
It occurs 
even when the {\it long-range\/} correlations vanish as they do
for interfaces in the 
KPZ dynamic universality class, to which the present models belong for
all finite, nonzero values of $H$. 
 Skewness 
has also been 
observed in several other SOS-type models, such as the body-centered SOS model 
studied by Neergaard and den~Nijs,\cite{NEER97} the model for step propagation 
on crystal surfaces with a kink-Ehrlich-Schwoebel barrier studied 
by Pierre-Louis et al.,\cite{PIER99} and a model for the local time 
horizon in parallel kinetic MC simulations studied by Korniss 
et al.\cite{KORN00C} No skewness was observed for the SOS model with the soft Glauber 
dynamics.\cite{RIKV02} However, a small skewness was observed for the Ising model (whose
 interfaces include bubbles and overhangs) with soft Glauber
dynamics (about two orders of magnitude smaller than the skewness observed for the hard 
Glauber dynamics).\cite{RIKV03}
The correlations associated with the skewness generally lead to a broadening 
of protrusions on the leading edge (``hilltops''), while 
those on the trailing edge (``valley bottoms'') are sharpened,\cite{NEER97} 
or the other way around.\cite{KORN00C} In terms of spin-class populations, 
the former corresponds to $\langle n(21-) \rangle > \langle n(21+) \rangle$ 
and $\langle n(11+) \rangle > \langle n(11-) \rangle$. The relative skewness 
can therefore be quantified by the two functions,\cite{NEER97} 
\begin{equation}
\rho = \frac{\langle n(21-) \rangle - \langle n(21+) \rangle}
{\langle n(21-) \rangle + \langle n(21+) \rangle}
\;,
\label{eq:rho}
\end{equation}
and\cite{RIKV02B}
\begin{equation}
\epsilon = \frac{\langle n(11+) \rangle - \langle n(11-) \rangle}
{\langle n(11+) \rangle + \langle n(11-) \rangle}
\;.
\label{eq:epsi}
\end{equation}
These two skewness parameters are shown together in Fig.~\ref{fig:skew}(a)
for the TDA, and in Fig.~\ref{fig:skew}(b) for the OSD.  
For both dynamics the relative skewness is seen to be considerably stronger at the lower 
temperature. The temperature dependence is especially pronounced for $\rho$,
 due to the low concentration of sites in the class 21+ at low temperatures. 
The OSD results contrast with previous results that showed that an SOS interface
with soft Glauber dynamics does not present skewness.\cite{RIKV02} In the
present case, the SOS interface evolving
under the soft OSD dynamics
presents stronger relative skewness at lower temperature than the one evolving under the hard TDA dynamics.

\begin{figure}[ht] 
\includegraphics[angle=0,width=.48\textwidth]{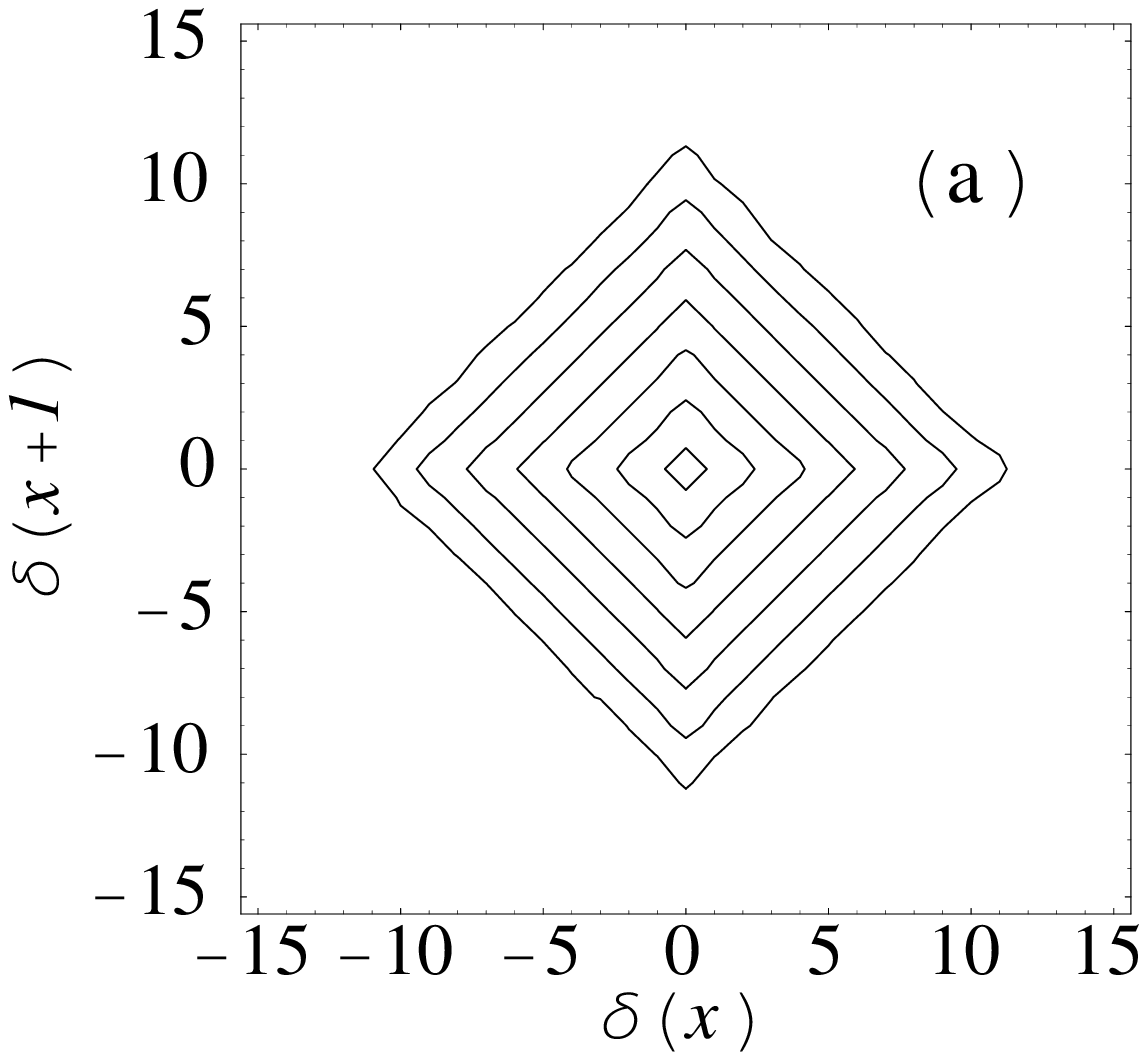} 
\includegraphics[angle=0,width=.48\textwidth]{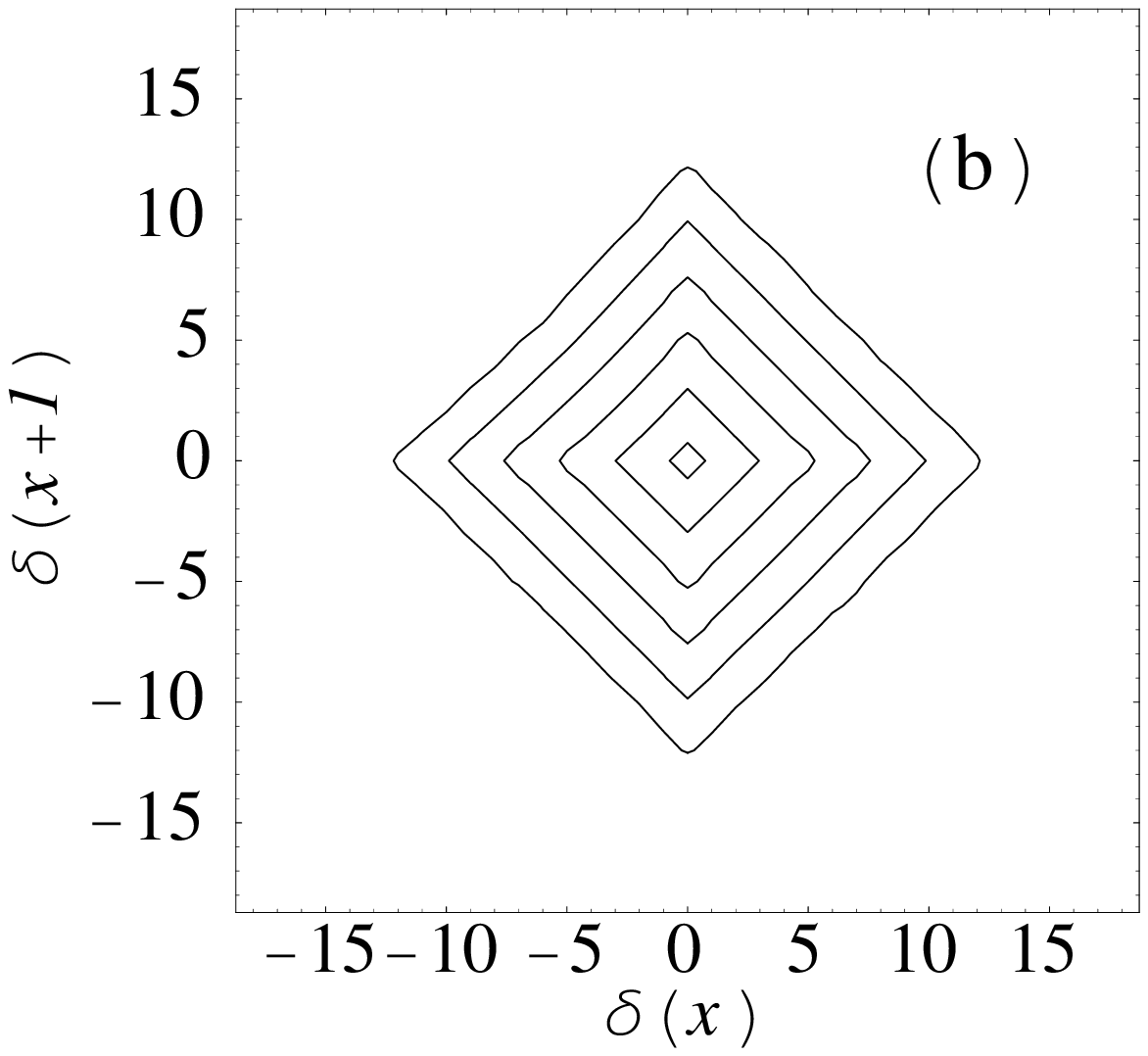} \\
\vspace{0.5truecm}
\includegraphics[angle=0,width=.48\textwidth]{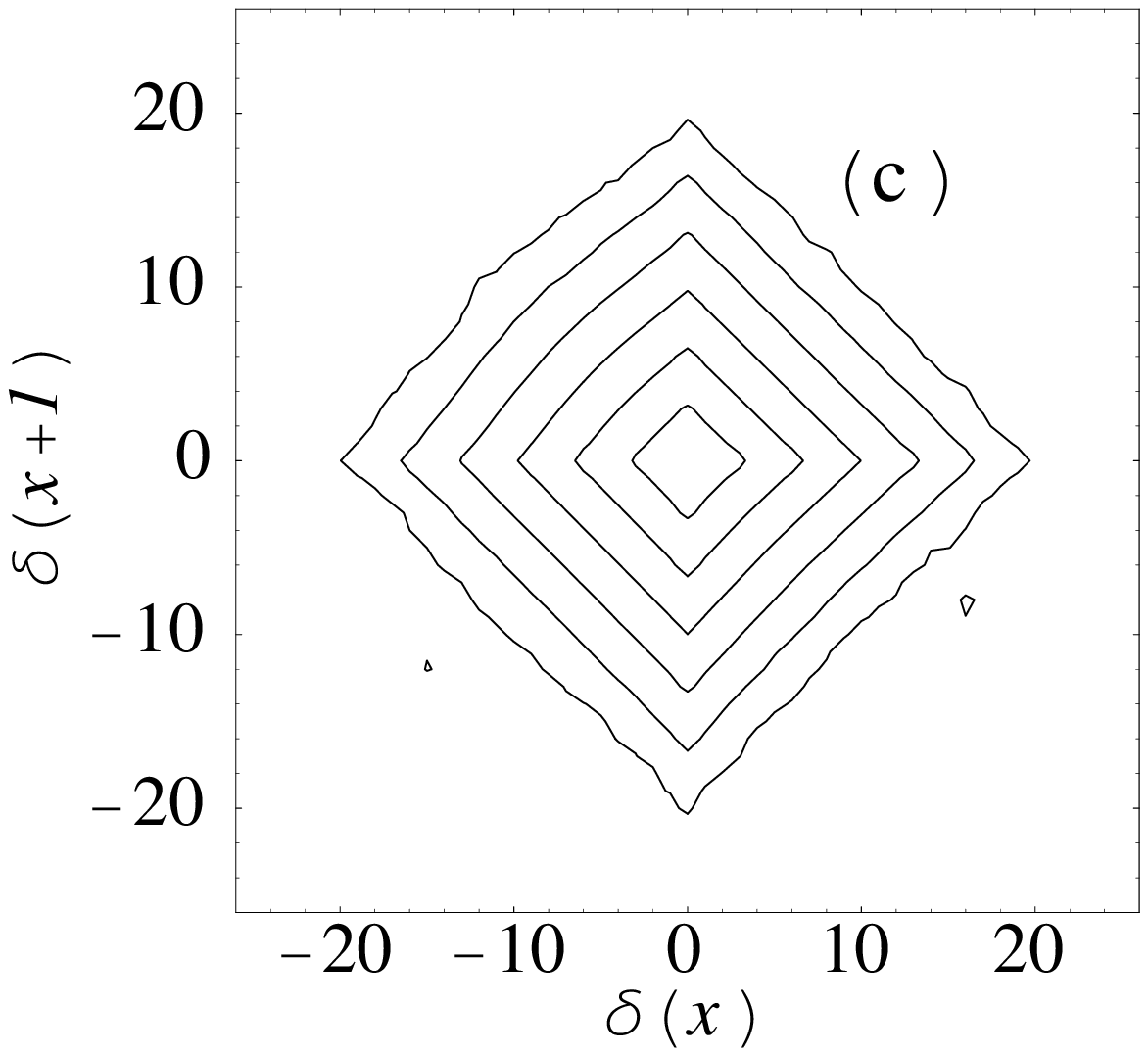}
\includegraphics[angle=0,width=.48\textwidth]{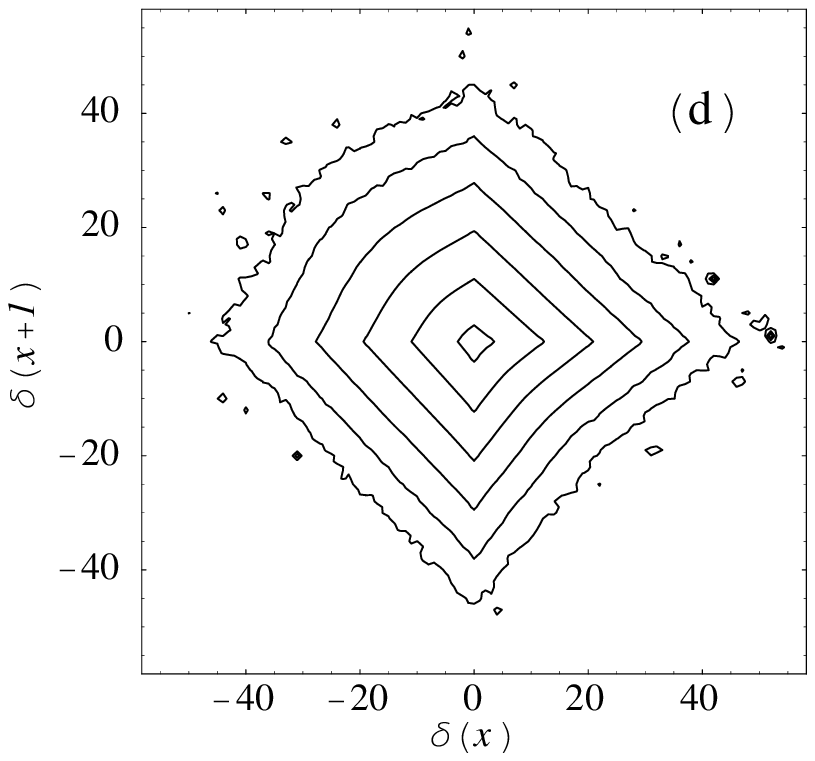}
\caption[]{
Contour plots of 
$\log_{10} p \left[ \delta(x) , \delta(x+1) \right]$ for $\phi=0$ at
$T=0.6T_c$ for the TDA dynamics. 
(a)
$H/J = 0$. 
(b)
$H/J = 1.0$. 
(c)
$H/J = 2.0$.
(d)
$H/J =3.5$.
Note the different scales in the four parts. 
See discussion in the text.  
}
\label{fig:cont_TDA}
\end{figure}

\begin{figure}[ht]
\includegraphics[angle=0,width=.48\textwidth]{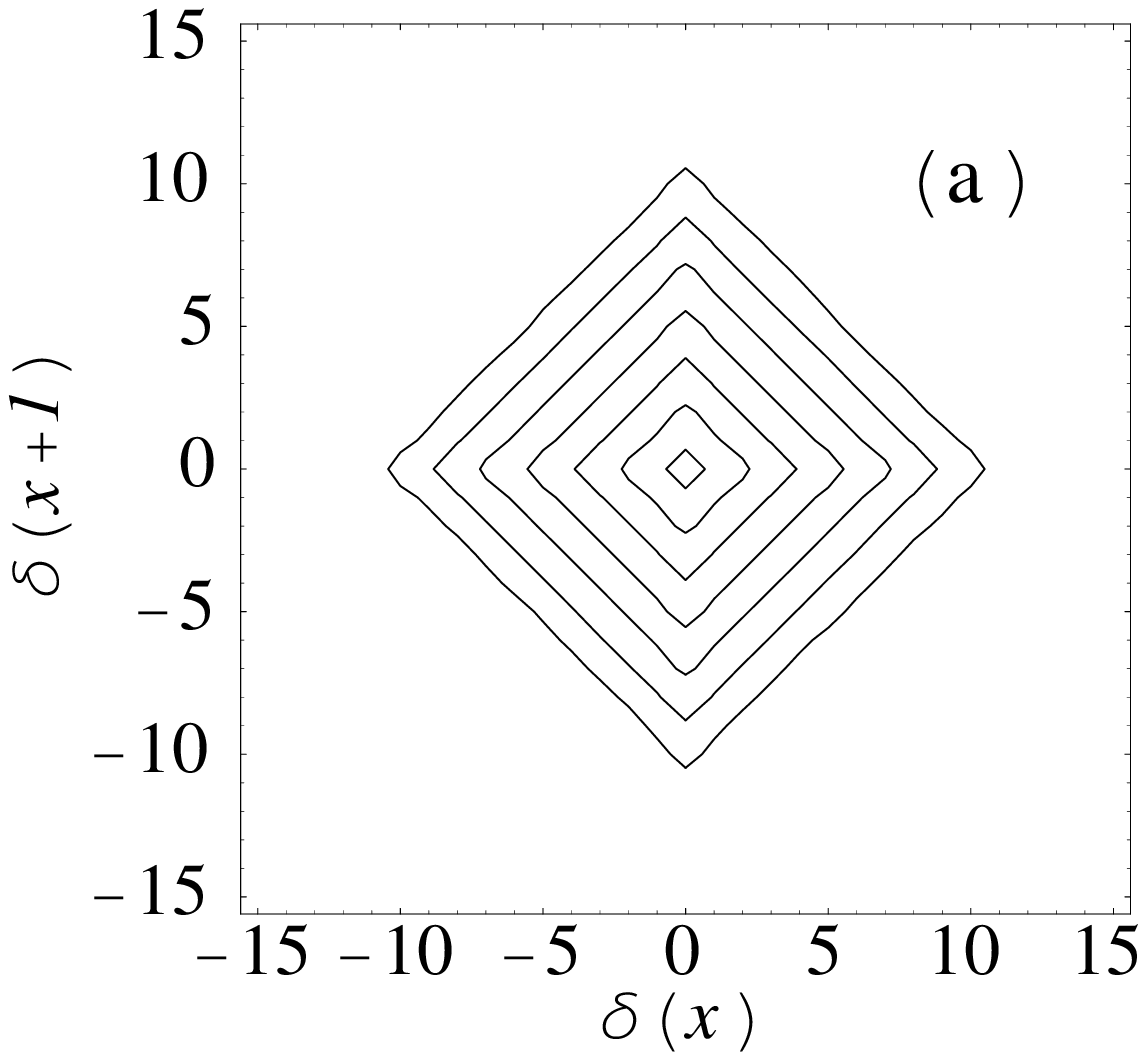}
\includegraphics[angle=0,width=.48\textwidth]{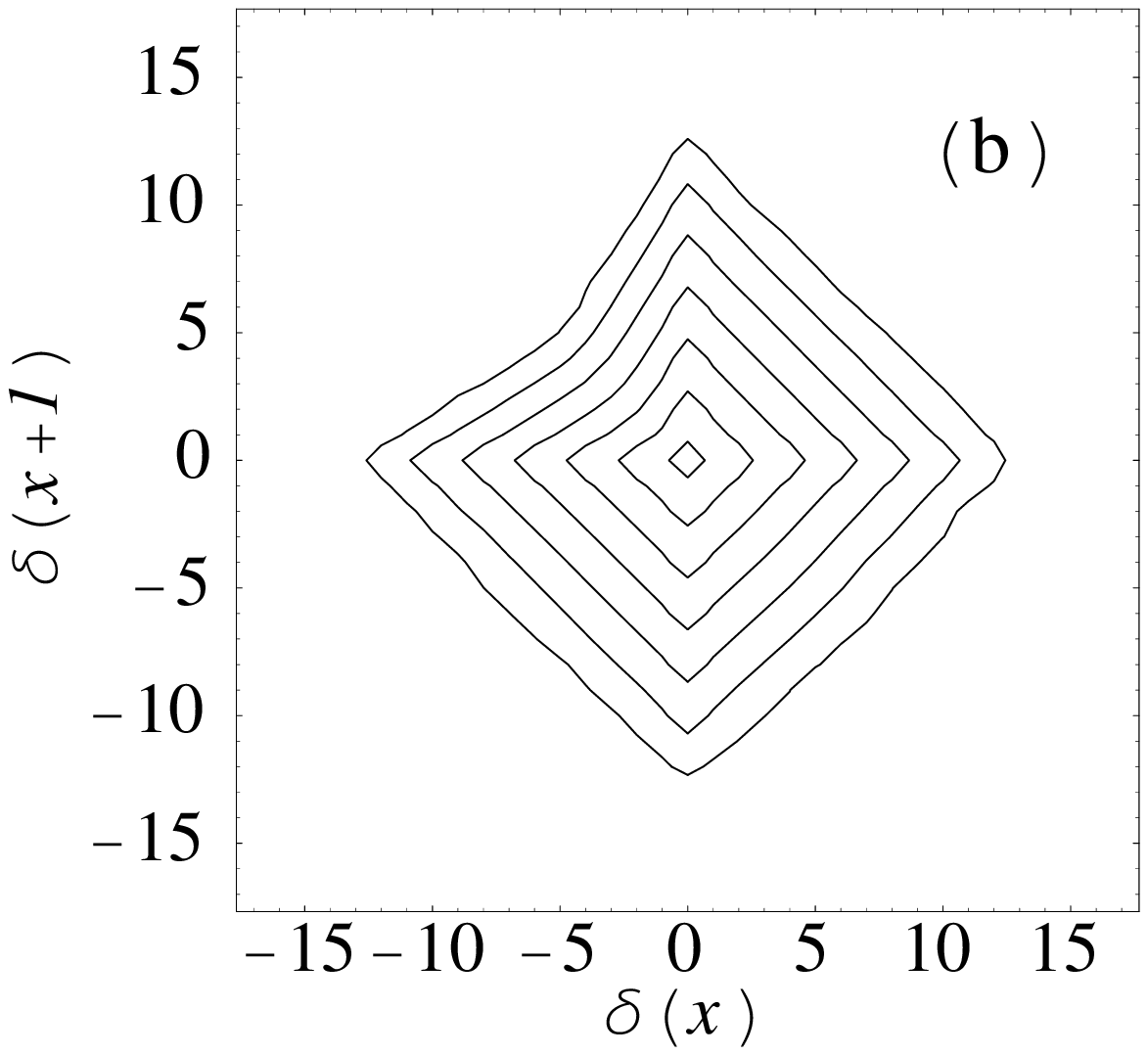} \\
\vspace{0.5truecm}
\includegraphics[angle=0,width=.48\textwidth]{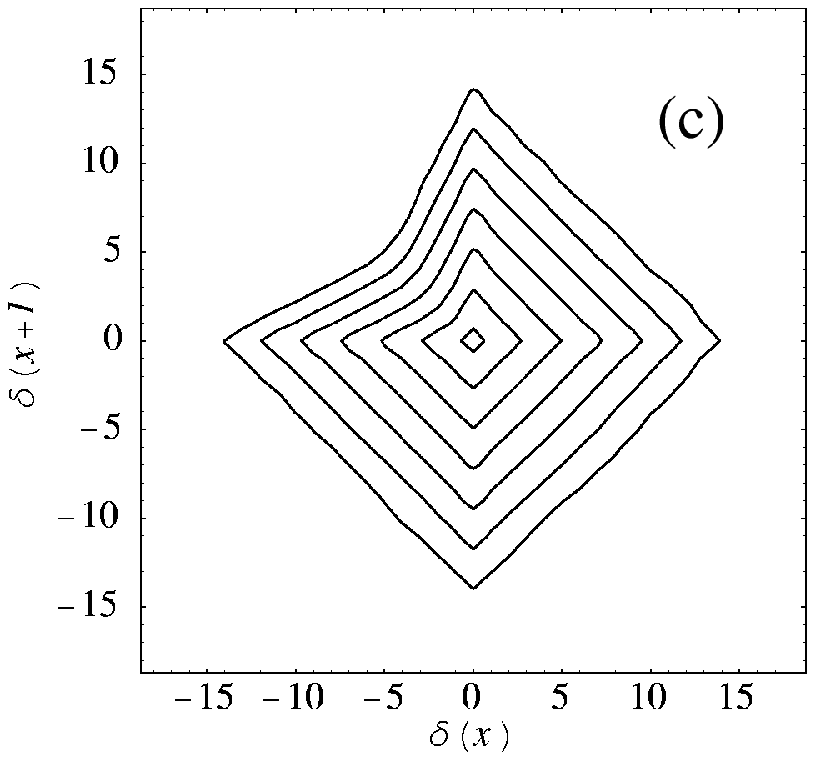}
\includegraphics[angle=0,width=.48\textwidth]{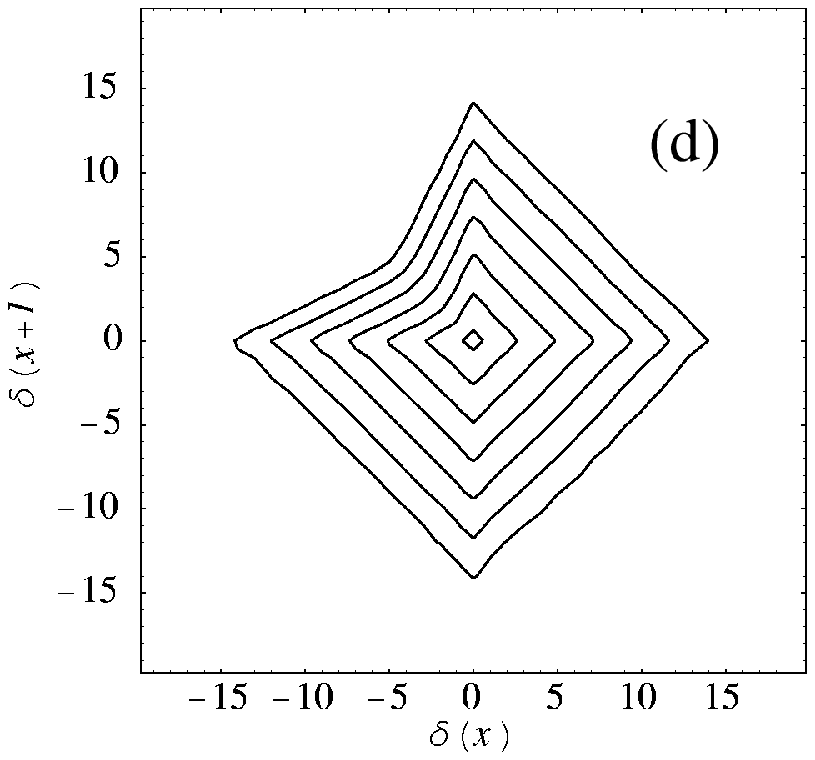}

\caption[]{
Contour plots of
$\log_{10} p \left[ \delta(x) , \delta(x+1) \right]$ for $\phi=0$ at
$T=0.6T_c$ for the OSD dynamics.
(a)
$H/J = 0$.
(b)
$H/J = 1.0$.
(c)
$H/J = 2.0$.
(d)
$H/J = 2.5$.
See discussion in the text.
}
\label{fig:cont_1SD}
\end{figure}

Another way to visualize the skewness is to consider the joint 
two-step pdf, $p \left[ \delta(x),\delta(x+1) \right]$. Logarithmic contour 
plots of this quantity for both dynamics at different values of $H$, for $\phi=0$
 at $T=0.6T_c$, 
are shown in Fig.~\ref{fig:cont_TDA} and Fig.~\ref{fig:cont_1SD}. It is clearly 
seen that in both dynamics the contours change with $H$. 
For $H$=0 a symmetric diamond shape with equidistant contours indicates 
statistical independence with single-step pdfs given by Eq.~(\ref{eq:step_pdf}). This equilibrium
 result 
is correctly observed in both the TDA dynamics, Fig.~\ref{fig:cont_TDA}(a), and 
the OSD dynamics, Fig.~\ref{fig:cont_1SD}(a).  
However, for nonzero fields the dynamics show different behavior. For the TDA,
at stronger fields, the shapes are convex in the second 
quadrant [$\delta(x) < 0, \; \delta(x+1) > 0$] 
 Fig.~\ref{fig:cont_TDA}(c) and (d). 
This shape indicates that large negative $\delta(x)$ tend to be followed by 
large positive $\delta(x+1)$ (sharp valleys). 
 For weak fields, the TDA shows a weak skewness of the 
opposite sign, as seen in Fig.~\ref{fig:cont_TDA}(b). For the OSD dynamics, for all 
nonzero fields, the shapes are 
concave in the second
quadrant, 
 Fig.~\ref{fig:cont_1SD}(b), (c), and (d).
This shape indicates that large negative $\delta(x)$ tend to be followed by
smaller positive $\delta(x+1)$ (rounded valleys).
For both dynamics the field dependence in the fourth quadrant, corresponding to the
shape of ``~hilltops,'' is much weaker that in the second quadrant. As expected, the contour plots for interfaces with $\phi=0$ are always symmetric
about the line $\delta(x+1) = - \delta(x)$ for both dynamics.

\section{Discussion and Conclusions}
\label{sec:DISC}

In this work we have continued our studies of the microstructure of an unrestricted 
SOS interface driven far from equilibrium by an applied field. 
 Previous studies indicate that different dynamics can lead to important differences in the microstructure of the
 moving interface and that extreme care therefore must 
be taken in selecting stochastic dynamics appropriate for the specific physical system of
interest.\cite{RIKV00B,RIKV02B,RIKV02,RIKV03}

 For this study we have considered two dynamics that include a local energy barrier 
representing a transition state inserted between the individual Ising or lattice-gas states. 
Such Arrhenius dynamics, as they are often called, are appropriate in kinetic MC simulations of discrete Ising or lattice-gas models in which the discrete states serve as
approximations for high-probability configurations in an underlying continuous potential.

The two Arrhenius dynamics that we considered are the commonly used one-step-dynamics (OSD)\cite{KANG89,FICH91} and the two-step transition dynamics approximation (TDA).\cite{ALAN92, ALAN92A, ALAN02} The OSD belongs
to the class of dynamics known as soft and the TDA to the class known as hard.\cite{MARR99}

We studied the microstructure and velocity of the SOS interface 
by kinetic MC simulations and by a non-linear
mean-field theory developed in previous papers.\cite{RIKV00B,RIKV02B} 
 We calculated the interface velocity as a function
of the driving field, temperature, and angle of the interface relative to the lattice axes. We also studied the
local shape of the interface in terms of the spin-class populations, the average height of 
a step, and the probability density
for individual steps in the interface. The theory predicts significant differences between interfaces moving under hard and soft dynamics.  For soft dynamics, interface 
structures should remain independent of the applied field, while there must be a clear 
dependence on the field for hard dynamics.  

For the TDA dynamics we found generally very good agreement between the theoretical predictions and the MC simulations. However, for the OSD dynamics we found that, contrary to the
 theoretical prediction, there is a weak but clear dependence of the interface structure on the field. This dependence is
manifest in the average stationary step height and in the mean stationary class populations, which both 
 show a weak dependence with the field that saturates for strong fields. As
a consequence of this dependence,
the theoretical results for the velocities do not match very closely the simulated results for low temperatures, particularly for strong fields and small angles.

The interfaces moving under the hard TDA dynamics present similar characteristics to 
those evolving under a hard Glauber dynamics. 
However, we found significant differences between the structure of surfaces evolving
under the soft OSD dynamics and the soft Glauber dynamics studied in Ref.~\onlinecite{RIKV02}. The velocities under the OSD
dynamics present a discontinuity at $T=0$ that is not observed for the soft Glauber dynamics.
More interesting is the existence of strong skewness in the 
OSD model. This indicates that lack of
skewness is not a necessary characteristic of soft dynamics, 
as earlier results seemed to suggest.\cite{RIKV02,RIKV03}

Within the mean-field approximation used here, individual steps of the
interface are assumed to be statistically independent. Short-range correlations are
not taken into account by this approximation. The skewness between the spin population
on the leading and trailing edges of the interface is a consequence of such
short-range correlations. For both the TDA and the OSD dynamics,
 the interfaces undergo a gradual breakdown of up/down symmetry for increasing
fields, which has also
been observed in 
other examples of driven interfaces.\cite{KORN00C,NEER97,PIER99} This breakdown is
clearly evident for the OSD model, and it is probably the reason
why the mean-field approximation misses the weak field dependence in the OSD interface structure.

It is obviously important to note and eventually to understand the discrepancies between the
theoretical mean-field predictions and the simulation results for the OSD dynamics. However,
on a qualitative level the theory predicts quite accurately the differences between the
structures generated by the two dynamics. The average step height for the TDA increases dramatically with increasing field,
 as accurately predicted by the theory. For the OSD, the step height does increase somewhat
 with $H$, in contrast to the theoretical prediction of $H$-independence, but the increase
is very small. In comparison with the TDA, the OSD surface
remains very smooth, with an average step height well below unity.

As in previous studies, our results indicate strong differences between interfaces moving under 
 different dynamics, emphasizing the need for extreme care in selecting the appropriate dynamics
for the physical system of interest. 
Even in cases where soft dynamics are the more 
appropriate choice, as for solidification or adsorption 
problems where the
driving force is a chemical-potential 
difference,\cite{GUO90,KOTR91,MITC00C,HAMA04,HAMA05} 
the results can depend significantly on which soft dynamics are chosen.

\section*{Acknowledgments}
\label{sec:ACK}

G.~M.~B.\ appreciates the hospitality of the 
School of Computational Science at Florida State University. 
This research was supported in part 
by National Science Foundation Grant Nos.~DMR-0240078 and DMR-0444051,
 by Florida State 
University through the Center for Materials Research and Technology and
the School of Computational Science, by  the National High Magnetic Field Laboratory, and
by the Deanship of Research and Development of Universidad Sim{\'o}n Bol{\'{\i}}var.




%
%
\begin{table}[ht]
\caption[]{
The spin classes in the anisotropic square-lattice SOS model. 
The first column contains the class labels, $jks$. 
The second column contains the total 
field and interaction energy for a spin in each class, $E(jks)$, relative 
to the energy of the state with all spins parallel and $H=0$, 
$E_0 = -2(J_x + J_y)$. 
The third column contains the change in the total system energy 
resulting from reversal of a spin from $s$ to $-s$, $\Delta E(jks)$. 
The fourth and fifth columns contain $E_T - E_i$ and $E_f - E_T$, respectively.
The first three classes have nonzero populations in the 
SOS model, and flipping a spin in any of them preserves the SOS configuration. 
The other two classes (marked  $\dag$) 
also have nonzero populations in the 
SOS model, but flipping a spin in any of them would produce an overhang 
or a bubble and is therefore forbidden. 
Not that in this table, $s$ represents the spin value {\it before\/}
the spin flip. 
}
\begin{tabular}{| l | l | l | l | l |}
\hline
Class, $jks$ 
& $E(jks) - E_0$ 
& $\Delta E(jks)$ 
& $E_T - E_i$
& $E_f - E_T$
\\ 
 \hline\hline
 $01s$ 
 & $-sH + 2J_y$ 
 & $2sH + 4J_x $
 & $sH + 2J_x + U$
 & $sH + 2J_x - U$
\\
 \hline
 $11s$  
 & $-sH + 2(J_x+J_y)$ 
 & $2sH $
 & $sH + U$
 & $sH - U$
\\
 \hline
 $21s$  
 & $-sH + 2(2J_x+J_y)$ 
 & $2sH - 4J_x $
 & $sH - 2J_x + U$
 & $sH - 2J_x - U$
\\
 \hline\hline
$10s$ $\dag$ 
 & $-sH + 2J_x$ 
 & $2sH + 4J_y $
 & $sH + 2J_y + U$
 & $sH + 2J_y - U$
\\
 \hline
$20s$  $\dag$
 & $-sH + 4J_x$ 
 & $2sH - 4(J_x-J_y) $
 & $sH - 2(J_x-J_y) + U$
 & $sH - 2(J_x-J_y) - U$

\\
\hline
 \end{tabular}
\label{table:class}
\end{table}

\begin{table}[ht]
\caption[]{
The mean populations for the spin classes of the SOS 
interface, with the corresponding contributions to the interface velocity 
under the TDA and the  
OSD dynamics. The first column contains the class labels, $jks$. 
The second column contains the mean spin-class populations for 
general tilt angle $\phi$, with $\cosh \gamma(\phi)$ from Eq.~(\ref{eq:chgam}). 
The third and fourth columns contain the contributions to the mean interface 
velocity in the $y$ direction from spins in classes $jk-$ and $jk+$, 
Eq.~(\protect\ref{eq:generalv}), using the SOS-preserving TDA and OSD dynamics respectively.
For the TDA dynamics, $X=X(T,H)$ is given by Eq.~(\ref{eq:XTDA}), and for the OSD dynamics
$X$ is independent of $H$, and is given by Eq.~(\ref{eq:XOSD}).
In the third column $ A=\left[ \cosh \left( 2 \beta J_x \right) \cosh \left( \beta H \right) +
      \cosh\beta U \right]$, and $B=\sinh\left(2 \beta J_x\right) \sinh\left( \beta H \right)$.
}

\begin{tabular}{| l | l | l | l | }
\hline
Class, $jks$ 
& $\langle n(jks) \rangle$ 
& $\langle v_y(jk) \rangle_{\rm TDA}$ 
& $\langle v_y(jk) \rangle_{\rm OSD}$

\\ 
 \hline\hline
 $01s$  
 & $\frac{1 - 2X \cosh \gamma (\phi) + X^2}{(1-X^2)^2}$ 
 & $ e^{-2 \beta J_x}\frac{A\sinh \left( \beta H \right) + 
          B \cosh \left( \beta H \right)}{A^2-B^2}$
 & $ e^{-\beta \left(U + 2J_x \right)}2\sinh\left(\beta H \right)$
                                                                                                            
\\
 \hline
 $11s$  
 & $\frac{2X[(1+X^2) \cosh \gamma (\phi) - 2X]}{(1-X^2)^2}$

 & $\frac{\sinh\left( \beta H \right)}{\cosh \left( \beta H \right) + \cosh \beta U}$

 & $e^{-\beta U}2\sinh\left(\beta H \right)$
\\
 \hline
 $21s$  
 & $\frac{X^2[1-2X\cosh\gamma(\phi)+X^2]}{(1-X^2)^2}$ 
 & $e^{2 \beta J_x}\frac{A \sinh \left( \beta H \right) - B\cosh \left( \beta H \right)}
               {A^2-B^2}$

 & $e^{-\beta \left( U -2J_x \right)}2\sinh\left(\beta H \right)$

\\
 \hline\hline
 \end{tabular}
\label{table:class2}
\end{table}
\clearpage 

\clearpage

\clearpage

\clearpage

\clearpage

\clearpage

\clearpage

\clearpage


\end{document}